\documentclass[11pt]{article}
\usepackage{fleqn,cospar}

\usepackage{url}

\usepackage{graphicx}
\usepackage[figuresright]{rotating}

\newcommand{\ms}{M$_{\odot}$}
\newcommand{\ls}{L$_{\odot}$}

\title{ HISTORY OF STAR FORMATION RATE AND LUMINOSITY DENSITY OF GALAXIES}

\author{T. N. Rengarajan and Y. D. Mayya\address{Instituto Nacional Astrof\'{i}sica \'{O}ptica y Electr\'{o}nica,
 Tonantzintla, Puebla 72840, Mexico}}
\begin{document}

\maketitle

\begin{abstract}
We have computed the time evolution of bolometric, far-infrared,
H$\alpha$ line, ultraviolet (both intrinsic and escaping the star
forming region) and the nonthermal radio continuum luminosities
for  continuous and constant star formation terminating at 95 Myr.
The luminosity rises to a plateau value and declines after the
termination of starburst, but only gradually. The time evolution
profiles are broad and different for different star formation
indicators. The broad profiles lead to uncertainties in the star
formation rate derived depending on the initial mass function,
duration of starburst, its distribution and the observational
epoch.

\end{abstract}

\section*{INTRODUCTION }
\vspace{\baselineskip}
 In recent years there have been several
studies of the cosmological history of star formation rate (SFR)
in galaxies. Usually the SFR is derived from the luminosity
density in a selected wavelength band which is an indicator of
star formation. Some of the SF indicators that have been used are:
H$\alpha$ line, far-infrared (FIR), submm, ultraviolet (UV) and
nonthermal radio continuum emissions. Each indicator is sensitive
to a certain range of stellar masses and the total SFR is obtained
by assuming an initial mass function (IMF) of stars at birth,
usually a power-law form. Further, the observed luminosity depends
on the star formation history and the epoch of observation. Even
for a constant rate of star formation, there is a time evolution
of luminosity which increases, reaches an equilibrium value and
declines after the termination of SF. This evolution and decline
also depend on the exponent of the IMF. The changes in the
luminosity to SFR ratio  as a function of these parameters are not
the same for all indicators. Thus, apart from the observational
errors, uncertainties arise in the SFR derived from luminosity
density using different indicators. In this paper, we study these
uncertainties by computing the evolution of the luminosity to SFR
ratio for different IMFs and scenarios of star formation. It
should be noted that there are other parameters like metallicity
that can affect the derivation of SFR from luminosity. For
example, Hirashita et al (2001) find that the FIR-SFR conversion
factor can vary by a factor of up to 4 on this account. When we
have better knowledge of factors like metallicity and cirrus
contribution in starburst galaxies, it may be possible to quantify
these uncertainties.

\section*{COMPUTATION }
\vspace{\baselineskip}
 In this study we compute the time evolution
of L/SFR ratio for the following luminosities: L$_{bol}$, the
bolometric luminosity, L$_{fir}$, the far-infrared luminosity
which is a fraction of the bolometric luminosity, L$_{uvi}$, the
intrinsic luminosity in an ultraviolet band, L$_{uve}$, the
luminosity of uv photons escaping the star forming region,
L$_{H}$, the H$\alpha$ line luminosity and L$_{R}$, the nonthermal
radio continuum luminosity. For UV we use the 0.16
 $\mu$m band which has been extensively used in HST observations.
The submm luminosity is usually converted into a FIR luminosity to
derive the SFR and hence we do not consider it separately. The
radio continuum is assumed to be from synchrotron emission from
cosmic ray electrons accelerated by supernovae arising from stars
of mass $>$ 8 \ms. In starburst(SB) galaxies, which are the
subjects of our study, the contribution of the earlier disc
population is small and is neglected. Their contribution, if
present, will add additional uncertainties. The FIR luminosity is
dust-reradiated emission and in many studies, it has been taken to
be the same as the bolometric luminosity. Though this is likely to
be the case at the very early embedded stage of a star, the
efficiency of conversion will be less than unity and decrease  as
time passes because of the disruption and dispersion of the parent
clouds by the massive stars formed. The fact that visible OB stars
are observed in  star forming complexes in the Milky Way supports
this hypothesis. Silva et al.\ (1998) have introduced the concept
of residence time and have shown that their models give good fits
to observational spectral energy distributions and spectral line
data. In this study, we assume that the FIR conversion efficiency
decreases exponentially with a time constant $\tau$. It then
follows that the flux of UV photons escaping the star forming
region  depends inversely on this conversion efficiency and can be
characterized by an extinction which decreases exponentially with
the same time constant. The H$\alpha$ luminosity will also have a
similar dependence with a proper scaling of the extinction.

The total luminosity in a given band is the integral over the
history of stellar formation and evolution and over the stellar
masses. The IMF is assumed to have a power law form $\xi$(m)
$\propto$ m$^{-\gamma}$ within mass limits of 120 to 1 \ms. Since
the contribution of stars of mass $<$ 1 \ms~  to the luminosity,
in the bands considered, is negligible, the L/SFR ratio can easily
be scaled for an IMF extending to lower masses. The stellar
evolutionary tracks are taken from Schaller et al.\ (1992) and the
computational procedure follows that of Mayya (1995). The
integrated luminosity at time T, the epoch of observation is given
by

\begin{equation}
L_{bol}(T) = \int_{1}^{120} \xi(m)dm \int_{T}^{0}p(t) L_{bol}(t,m)
dt
\end{equation}

\begin{equation}
L_{fir}(T) = \int_{1}^{120} \xi(m)dm \int_{T}^{0}p(t) L_{bol}(t,m)
e^{-t/\tau} dt
\end{equation}

\begin{equation}
L_{uvi}(T) = \int_{1}^{120} \xi(m)dm \int_{T}^{0}p(t) L_{uv}(t,m)
 dt
\end{equation}

\begin{equation}
L_{uve}(T) = \int_{1}^{120} \xi(m)dm \int_{T}^{0}p(t) L_{uv}(t,m)
10^{-0.4 A(uv)exp(-t/\tau)} dt
\end{equation}

\begin{equation}
L_{H}(T) = K_{H} \int_{1}^{120} \xi(m)dm \int_{T}^{0}p(t)
L_{Lyc}(t,m) 10^{-0.4 A(H)exp(-t/\tau)} dt
\end{equation}

\begin{equation}
L_{R}(T) = K_{R } \int_{8}^{120} \xi(m)dm \int_{T}^{Min(T,
t^{*}(m))}p(t) e^{-t/\tau_{e}} dt
\end{equation}

Here t is the look back time when a star is born, T is the
observational epoch from a reference time when the starburst
started, p(t) is the production rate of stars at time t,
t$^{*}$(m) is the lifetime of a star of mass m, A(uv) and A(H) are
the extinctions in magnitudes at 0.16 $\mu$m and H$\alpha$
wavelength at the time of birth, Lyc refers to Lyman continuum
photons and K$_{H}$ is a constant that relates H$\alpha$
luminosity to the Lyman continuum luminosity and K$_{R}$ is the
nonthermal radio continuum luminosity per supernova and $\tau_{e}$
is the exponential residence time of synchrotron emitting
electrons. The production term p(t) represents the time evolution
of the starburst. It is a $\delta$ function for an instantaneous
burst, constant for a continuous and constant star formation and
could have other forms like an exponential decay or a gaussian.

The IMF index is neither known accurately nor is it known to be
universal. Observations of stellar clusters in the Milky Way and
the Magellanic clouds give a range of 2 to 3 ( Scalo 1998). We,
therefore, perform computations for three values of $\gamma$ = 2,
2.5 and 3. As for the time dependence of star formation, we will
consider, in this paper, only the case of p(t) = constant for t =
0 to T$_{B}$ and zero elsewhere and present the results for
$T_{B}$ = 95 Myr. As for the residence time of stars in the parent
clouds, Silva et al.\ (1998) find that for the starburst galaxies
they model, a value of 10--20 Myr fits the observations. Here, we
take $\tau$ = 15 Myr. The extinction at 0.16 $\mu$m and H$\alpha$
are taken as 10 and 3 magnitudes respectively. The UV extinction
adopted leads to a luminosity weighted  extinction factor of 5-6
at times $>$ 50 Myr similar to the value estimated by  Steidel et
al.\ (1999). It may be noted that for moderately obscured regions,
a more elaborate attenuation model is needed. However, for the
luminous starburst galaxies, especially at moderate and large
redshifts, our treatment may be adequate. As remarked earlier the
presence of older star-forming regions will add further
uncertainty to the SFR. For the residence time of high energy
electrons, we take $\tau_{e}$ = 10 Myr.

\section{RESULTS AND  DISCUSSION}

\vspace{\baselineskip}
 In Figure 1, we show for different SF indicators,
the time evolution of the ratio of luminosity at time T to that at
95 Myr when the SB terminates. We have not shown the curve for
H$\alpha$ since it is very similar to that of the FIR. For FIR and
UVE, we show the ratio with respect to the bolometric and
intrinsic uv luminosity respectively. The general trend of the
time evolution is an increase of the luminosity to a plateau value
and then a decrease after the SB terminates.The plateau values of
the ratio of luminosity to SFR for the bolometric, FIR and uv
(0.16 $\mu$m)emissions are 1.6$\times$10$^{10}$ \ls \ms$^{-1}$ Yr,
8.8$\times$10$^{9}$ \ls \ms$^{-1}$ Yr and 5.5$\times$10$^{20}$ W
Hz$^{-1}$ \ms$^{-1}$ Yr respectively. These values, after
adjustment for the Salpeter IMF and the 0.1--100 \ms~ range
adopted by Kennicutt ( 1998),are about the same as those given in
his review. Further, the ratio of FIR to radio continuum fluxes is
similar to that of Bressan et al (2002) if we adopt their value of
nonthermal emission per supernova.

\begin{figure}[t]
\begin{minipage}{120mm}

\includegraphics[width=100mm]{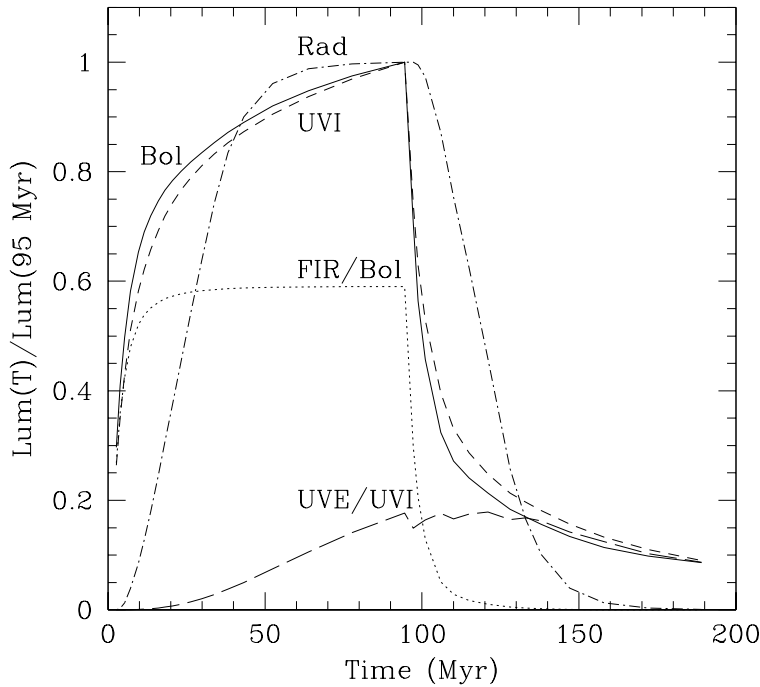}
\vspace{-7mm}
 \caption{Time evolution of luminosities for a continuous and constant
 star formation and IMF index $\gamma$ = 2.5. For FIR and UVE, escaping
uv luminosity, ratio with respect to bolometric and intrinsic uv
luminosity respectively are shown with the same numerical scale.}
\end{minipage}
\end{figure}

\begin{figure}[t]

\begin{minipage}{120mm}

\includegraphics[width=100mm]{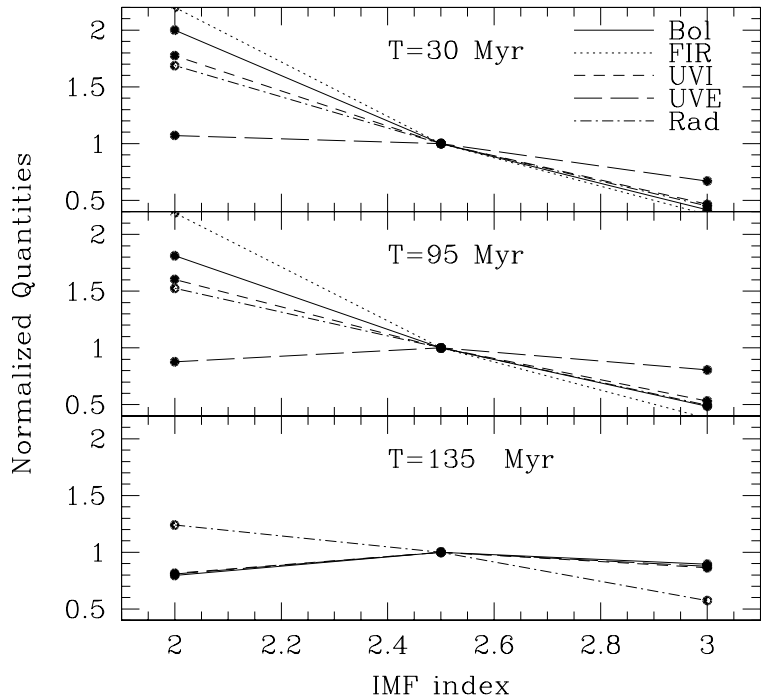}
\vspace{-7mm}
 \caption{Luminosity to SFR ratio
normalised to the value for IMF index $\gamma$ = 2.5 plotted
against $\gamma$ for three observational epochs. The points are
joined together to facilitate viewing.}
\end{minipage}
\end{figure}

  The rise- and fall-times vary from band to band. The
time profile for the FIR luminosity is the sharpest while it is
the broadest for the UVE. This is due to the fact that luminous
massive stars have lifetimes $<<$ the residence time and almost
the total bolometric luminosity is re-radiated in the FIR while
for stars with lifetime $>$ 15 Myr, the conversion efficiency
decreases. The opposite is the case for the escaping uv photons.
 For $\tau$ = 15 Myr, the
maximum conversion efficiency is 0.55. It will be lower if $\tau$
is less. The fraction of escaping uv photons reaches a maximum of
0.17. The uv extinction is more sensitive to the values of $\tau$
and extinction at zero age than the FIR conversion efficiency. The
slight increase in the UVE luminosity after the termination of the
SB and the subsequent oscillation is due to a competition between
the death of massive stars and increased escape of uv photons from
lower mass stars as the stars age. It may be noted that the FIR
profile is sharper than that of the bolometric luminosity while
the UVE approaches that of UVI in the post starburst period. The
radio continuum luminosity has a rise- and fall-time of about 30
Myr, the lifetime of stars of mass 8 \ms.  The plateau FIR and UVE
luminosities depend on $\tau$, the residence time. Hence, there
will be additional uncertainties in the SFR  resulting from the
variation of $\tau$ from galaxy to galaxy and its possible
dependence on the intensity of starburst. It is generally believed
that the FIR emission is least subject to extinction correction
applicable to other wavelengths. However, the concept of residence
time introduces corrections up to a factor of 2--3. From this
point of view, the radio continuum emission is a better probe of
SFR since it needs no such correction.

It is clear that the SFR derived for an individual galaxy depends
on the duration of the SB and the time of observation and the
luminosity does not give the instantaneous current SFR unless one
observes in the plateau region of continuous and constant star
formation.  Even if we observe a large enough  sample of galaxies,
say in a given redshift bin, and all galaxies have the same
duration of starburst and other parameters remain the same, the
uncertainty is not fully eliminated. If we assume that the SB
duration is the same for all galaxies, at any given time one will
observe galaxies distributed uniformly over the evolutionary
stage. Though the plateau luminosity may be higher than the
observational threshold, in the rising portion of the evolutionary
curve, galaxies will be missed when the luminosity is below the
threshold. In the post starburst period, one will observe galaxies
above the threshold though there is no ongoing star formation. For
example, if the luminosity is n times the observational threshold
and t$_{1}$ and t$_{2}$ are the times when the fraction of plateau
luminosity is 1/n in the rising and falling stages respectively (
see Fig. 1), the fraction of galaxies that will be missed is
t$_{1}$/T$_{B}$ and the additional contribution from the post SB
period is (t$_{2}$-T$_{B}$)/$T_{B}$. The effect depends on the SF
indicator used, the duration of the SB and the luminosity function
in the chosen band. The FIR emission is a good tracer of
'instantaneous' SFR since both the rise and fall are steep. For
the radio continuum, there is a net over-estimation since the
decline in the post SB period is slower. Using the 1.4 GHz radio
luminosity function of Haarsma et al (2000) for convolution and
data of Fig.1, we find the net over-estimation of the luminosity
density to be 26 \%. The over-estimation in the case of uv photons
is much higher. The corrections also depend on the duration of the
SB. For low values of SB duration, the plateau region is a small
fraction of the total and hence the correction is more. Thus, the
uncertainties are significant as long as the SB duration is $<$
150-200 Myr.

\subsection{Effect of IMF Variation}
 In Figure 2 we have plotted,  the
luminosity to SFR ratio normalized to the value for the IMF index
$\gamma$ = 2.5 vs $\gamma$ for three epochs, 30 Myr, 95 Myr and
135 Myr. It is seen that the variation with index is not the same
for the different bands. The FIR luminosity shows the maximum
variation, a factor of more than 5 as the index changes from 2 to
3. The escaping uv photons show the least change. Further, the
changes also depend on the observational epoch. Thus, for
individual galaxies, one can have large uncertainties in SFR if
the IMF varies. If a large sample of galaxies is available, the
average over varying index will result in reduced uncertainty.

It is not known whether the IMF extends from high mass to low mass
stars in the same region at the same time. In intense starbursts
it is believed that the  IMF is biased towards high mass stars. In
the Galactic star forming region W3, Megeath et al (1996) find
that on extrapolation of the observed IMF for M $<$ 10 \ms, they
expect to see only one star of mass  $>$ 10 \ms, but radio
continuum observations show the presence of six high mass stars.
If in starbursts, the IMF does not extend below, say 10 \ms, the
luminosity computed would not change much, but the total stellar
mass would be lower. If IMF bias is present and varies from galaxy
to galaxy, there will be additional uncertainties in the derived
SFR.

\section{SUMMARY}
\vspace{\baselineskip}
 We have computed the time evolution of
luminosity in different bands of star formation indicators for the
case of a continuous and constant star formation.  The luminosity
rises from the time the starburst begins, reaches a plateau value
and declines after the termination of the starburst. The
post-burst decline is generally, slower than the growth at early
times. We have adopted the concept of residence time to calculate
the FIR and escaping uv luminosities. The FIR luminosity has the
sharpest profile while the uv luminosity has the broadest. The
star formation rate derived from luminosity density is subject to
uncertainties depending on the shape of the IMF, the residence
time parameter, the duration and distribution of starburst and the
observational epoch.

\vspace{8mm}
 Email: renga@inaoep.mx (T. N. Rengarajan); ydm@inaoep.mx  ( Y. D.
 Mayya)\\

\noindent \vspace{8mm} Manuscript received: 28 October 2002;
revised: 13 May 2003; accepted: 14 May 2003
\end{document}